\documentclass[twocolumn,amsmath,amssymb,superscriptaddress]{revtex4}
\usepackage{graphicx}
\usepackage{dcolumn}
\usepackage{bm}
\usepackage{color}
\usepackage{wasysym}
\usepackage{tikz}

\DeclareMathOperator{\erfc}{erfc}

\renewcommand\vec{\mathbf}
\renewcommand{\(}{\left(}
\renewcommand{\)}{\right)}

\def \equi#1{\mathrel{\mathop{\kern 0pt\sim}\limits_{#1}}} 

\begin{document}

\title{Bath-Mediated Interactions between Driven Tracers
in Dense Single-Files}

\date{\today}
\author{Alexis Poncet}
\affiliation{LPTMC, CNRS/Sorbonne Universit\'e, 4 Place Jussieu, F-75005 Paris, 
France}
\author{Olivier B\'enichou}
\affiliation{LPTMC, CNRS/Sorbonne Universit\'e, 4 Place Jussieu, F-75005 Paris, 
France}
\author{Vincent D\'emery}
\affiliation{Gulliver, CNRS, ESPCI Paris, PSL Research University, 10 Rue Vauquelin, F-75005 Paris, France}
\affiliation{Laboratoire de Physique, ENS de Lyon, Université Lyon, Université Claude Bernard Lyon 1, CNRS, F-69342 Lyon, France}
\author{Gleb Oshanin}
\affiliation{LPTMC, CNRS/Sorbonne Universit\'e, 4 Place Jussieu, F-75005 
Paris, France}

\begin{abstract}
Single-file transport, where particles cannot bypass each other, has been observed in
various experimental setups.
In such systems, the behaviour of  a tracer particle (TP) is subdiffusive,
which originates from strong correlations between particles.
 These correlations are especially marked when the TP 
is driven and leads  to inhomogeneous density profiles.
Determining the impact of this inhomogeneity when {\it several} TPs are driven in the system is a key question, related to the general issue of bath-mediated interactions, which are known to induce collective motion and lead to the formation of clusters
or lanes
in a variety of systems. Quantifying this collective behaviour, the emerging interactions and their dependence on 
the amplitude of forces driving the TPs, remains a challenging but largely unresolved issue. Here, considering dense single-file systems, we analytically determine the entire dynamics of the  correlations and reveal out of equilibrium cooperativity and competition effects between driven TPs.
\end{abstract}

\maketitle

The motion of particles in narrow channels in which particles cannot bypass each other is
known as single-file diffusion.
Such systems have been studied
in various experimental setups with zeolites~\cite{Gupta:1995, Hahn:1996}, micro-~\cite{Wei:2000,Lin:2005} and nano-channels~\cite{Meersmann:2000}, and simulations of carbon nanotubes~\cite{Sasha:2002}.
Key features, on the theoretical side, involve the existence of 
a subdiffusive scaling
for the mean position of a given particle~\cite{Harris:1965,Arratia:1983}, and strong correlations between particles~\cite{Poncet:2018}.

The basic phenomenology of the single-file transport is well-captured by the symmetric exclusion
 process (SEP).
In this paradigmatic model of crowded equilibrium systems,
particles perform symmetric random walks 
on a one dimensional lattice with the constraint of at most a single occupancy of each lattice site.
Different facets of the SEP have been scrutinised
(see Refs.~\cite{Levitt:1973,Fedders:1978,Alexander:1978,Arratia:1983,Lizana:2010,Taloni:2008,Gradenigo:2012}),  
including several important extensions to out-of-equilibrium situations. 
In particular, the mean displacement~\cite{Einstein}, as well as all higher-order cumulants
~\cite{Imamura:2017} of an unbiased tagged particle (TP) placed initially at the shock point 
of a step-like density profile have been determined.
Moreover, for a SEP with a {\it single} biased  TP (due to either an energy consumption or an external force), the mean displacement of the latter ~\cite{Burlatsky:1996b,Landim:1998a} and the higher-order cumulants in the dense limit~\cite{Illien:2013} have been calculated, 
and shown to grow sublinearly as $\sqrt{t}$.
Here, 
the particles accumulate in front of the TP and are depleted behind it, which results in an inhomogeneous, non-stationary spatial distribution of particles.
 
A general open question concerns situations when {\it several} biased TPs are introduced in an otherwise quiescent medium of bath particles.
The TPs are then expected to entrain the bath particles in a directional motion, which brings the system out-of-equilibrium and gives rise to effective bath-mediated interactions (BMIs) between the TPs.  
Such BMIs potentially lead to self-organisation, as observed in systems 
as diverse as colloidal solutions \cite{Reichhardt:2006,Mejia-Monasterio:2011,Ladadwa:2013,Vasilyev:2017,Leunissen:2005,Rex:2008,Vissers:2011,Glanz:2012,Poncet:2017,Bain:2017}, nearly-critical fluid mixtures \cite{Furukawa:2013}, 
dusty complex plasmas \cite{Sutterlin:2009}
pedestrian counter flows \cite{Helbing:2001}.  

Quantifying the emerging interactions between  biased TPs
and the ensuing collective behaviour is thus a key issue which however remains largely unresolved.
Here, modeling a host medium as a dense SEP, 
we  analytically determine the temporal evolution of all
correlation functions,  
reveal intrinsically out-of-equilibrium cooperativity and competition effects between multiple TPs and quantify BMIs.

\begin{figure}
\begin{center}
\includegraphics{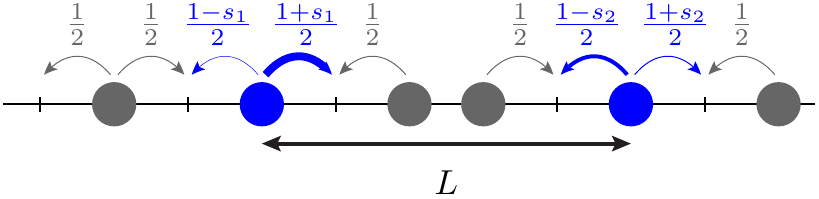}
\end{center}
\caption{System with two tagged particles (TPs). All particles perform random walks with unit jump rate, constrained by hard-core exclusion. Bath particles jump to the left or right with probability $1/2$. The TPs jump with probabilities $(1\pm s_j)/2$. $L$
is the initial distance between TPs.}
\label{fig:syst}
\end{figure}

The quiescent host medium is modelled as a SEP which involves
a high density $\rho$ {($\rho \to 1$)} of hard-core bath particles performing  {\it symmetric} random walks (with unit jump rate) on a one-dimensional lattice.
We then tag $N$ particles at initial positions $X_j^0$ (see Figure \ref{fig:syst}).
These TPs are biased: the $j$-th TP
jumps to the left (resp. right) with probability $(1-s_j) / 2$ (resp. $(1+s_j)/2$).
The bias $s_j \in (-1, 1)$ may either be due to an ``activity'' of the particle or
to an external force $f_j$, in which case one has 
the detailed balance condition: $e^{\beta f_j}=\frac{1+s_j}{1-s_j}$, $\beta$ being the reciprocal temperature.

We aim at determining the correlations between the TPs, embodied in the so called   cumulant-generating function $\psi(\vec k, t) \equiv \ln\left\langle e^{i \vec k \cdot \vec Y(t)}\right\rangle$, where $\vec k=(k_1, \dots ,k_N)$, $\vec Y=(Y_1, \dots ,Y_N)$ and $Y_j(t) = X_j(t) - X_j^0$ is the displacement of the $j$-th TP. The cumulants are denoted by $\langle\bullet\rangle_c$ and defined by the expansion
\begin{gather}
 \psi(\vec k, t)  = \sum_{p_1, \dots, p_N=0}^\infty
 \frac{(ik_1)^{p_1}\dots(ik_N)^{p_N}}{(p_1 + \dots + p_N)!}
 \left\langle Y_1^{p_1}\dots Y_N^{p_N}\right\rangle_c.
\end{gather}

In densely populated single-files ($\rho\to 1$), the dynamics of the system can be reformulated in terms of independent vacancies. In essence, this amounts to neglecting events where two vacancies interact simultaneously with any TP~\cite{Brummelhuis:1988a,Illien:2013}.
Our approach is based first on
considering an auxiliary problem involving a single vacancy, initially at position $U$. 
By counting all the interactions of this vacancy with the TPs, 
we determine the probability $p_U(\vec Y, t)$ that the TPs have displacements $\vec Y$ at time $t$. Then, 
for a density of vacancies $\rho_0 = 1 - \rho\to0$, the cumulant-generating function reads~\cite{SupplMat}
\begin{equation}
 \lim_{\rho_0\to 0} \frac{\psi(\vec k, t)}{\rho_0} =
 \sum_{U\notin\{X_i^0\}} [\tilde p_U(\vec k, t) - 1]
\end{equation}
where $\tilde p_U(\vec k, t)$ is the Fourier transform of $p_U(\vec Y, t)$ and can be expressed in terms of first-passage quantities of simple random  walks
with or without absorbing sites \cite{Hughes:1995}. 
Analysis of the explicit expression of the  cumulant-generating function for densely populated single-files~\cite{SupplMat} allows us to draw a number of important conclusions, which we present below.

{\it Bath-mediated binding.}  As expected, at short times, the TPs move independently subject to their own biases. Our first finding is that, at large time and at high particle density, they are moving as a single TP. More precisely, in the large-time limit the $N$-TPs cumulants 
are given by $\langle Y_1^{q_1}\dots Y_N^{q_N}\rangle_c = \langle Z^{q_1 +\dots+q_N}\rangle_c$ for positive integer $q_j$, 
where $Z=\sum_{j=1}^N Y_j/N$ is the displacement of
the center of mass.  At high density, even and odd cumulants of $Z$  satisfy
\begin{align} \label{eq:1tp}
 \frac{\langle Z(t)^{2n}\rangle_c}{\rho_0} &= \frac{\langle Z(t)^{2n+1}\rangle_c}{\rho_0 S}
 = \sqrt\frac{2t}{\pi},
\end{align}
where  $S=\tanh(\beta F/2)$ is
the effective bias, $F = \sum_{j=1}^N f_j$  the effective force and 
the ratio $\langle\bullet\rangle/\rho_0$ is  understood as
the limit $\rho_0\to 0$.
Equation \eqref{eq:1tp} implies in particular that for any number of  TPs and arbitrary forces  $\langle Y_j\rangle=\langle Z\rangle$, meaning that at large time all the TPs move like their center of mass with an effective force $F$ (in agreement with the hydrodynamic analysis of Ref.~\cite{Benichou:2018}).

In the following, we determine the full dynamics of the correlations between TPs, and focus for simplicity reasons on the case of two TPs.

\begin{figure}
\begin{center}
\includegraphics{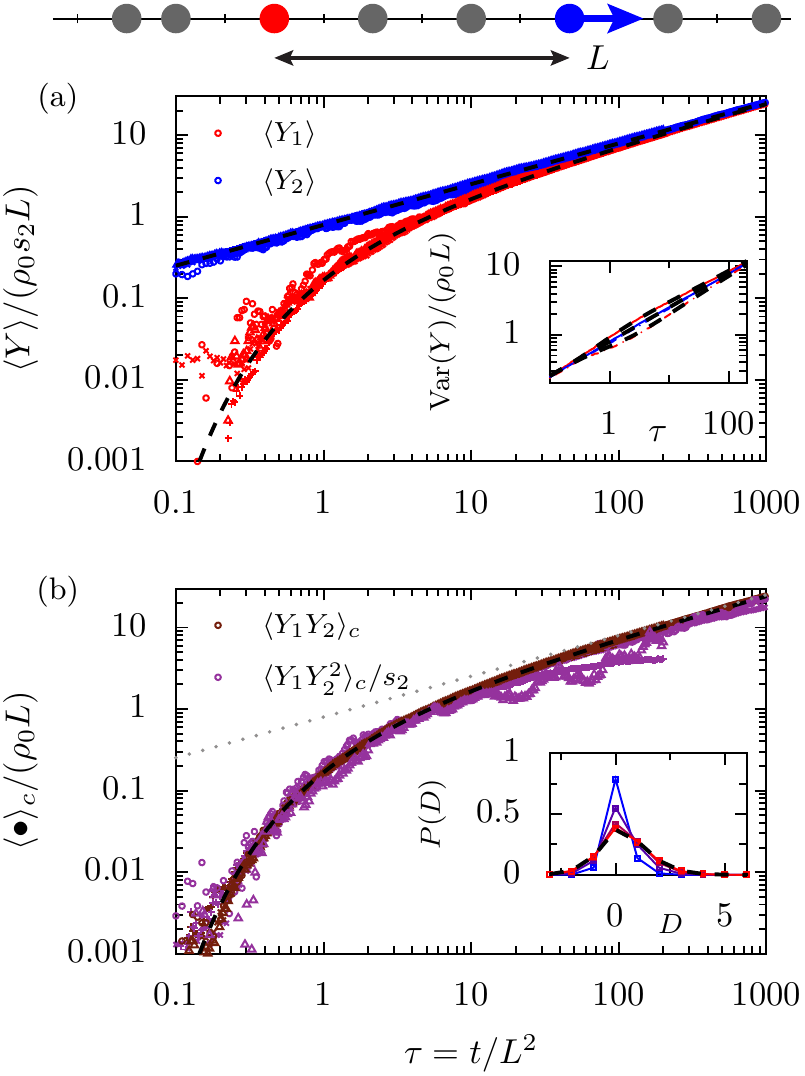}
\end{center}
\caption{
Entrainment of the bath particles by a biased TP for $\rho_0 = 10^{-2}$.  Only the right TP is biased. (a) Evolution of the mean displacements, different symbols corresponding  to $L=10, \;50$, $s_2=-0.2,\;0.8$. The black lines are the predictions from Eq.~\eqref{eq:displ1bias}.
(b) Cumulants $\langle Y_1Y_2\rangle_c$ and $\langle Y_1Y_2^2\rangle_c/s_2$ for the same set of parameters. The black line corresponds to Eq.~\eqref{eq:cums1bias}.
Inset of (a): variances for $L=10$, $s_2 = 0.8, -0.8$ (solid, dashes). Predictions in black~\cite{SupplMat}.
Inset of (b): law of the variation of distance $D=Y_2-Y_1$ at times $10, 10^2, 10^3, 10^4$ (blue to red) for $L=10, s=0.8, \rho_0=0.05$. The squares are the numerical results, the colored lines are the theoretical predictions, the black line is the asymptotic prediction~\cite{SupplMat}.
}
\label{fig:oneBias}
\end{figure}
\begin{figure*}
\begin{center}
\includegraphics{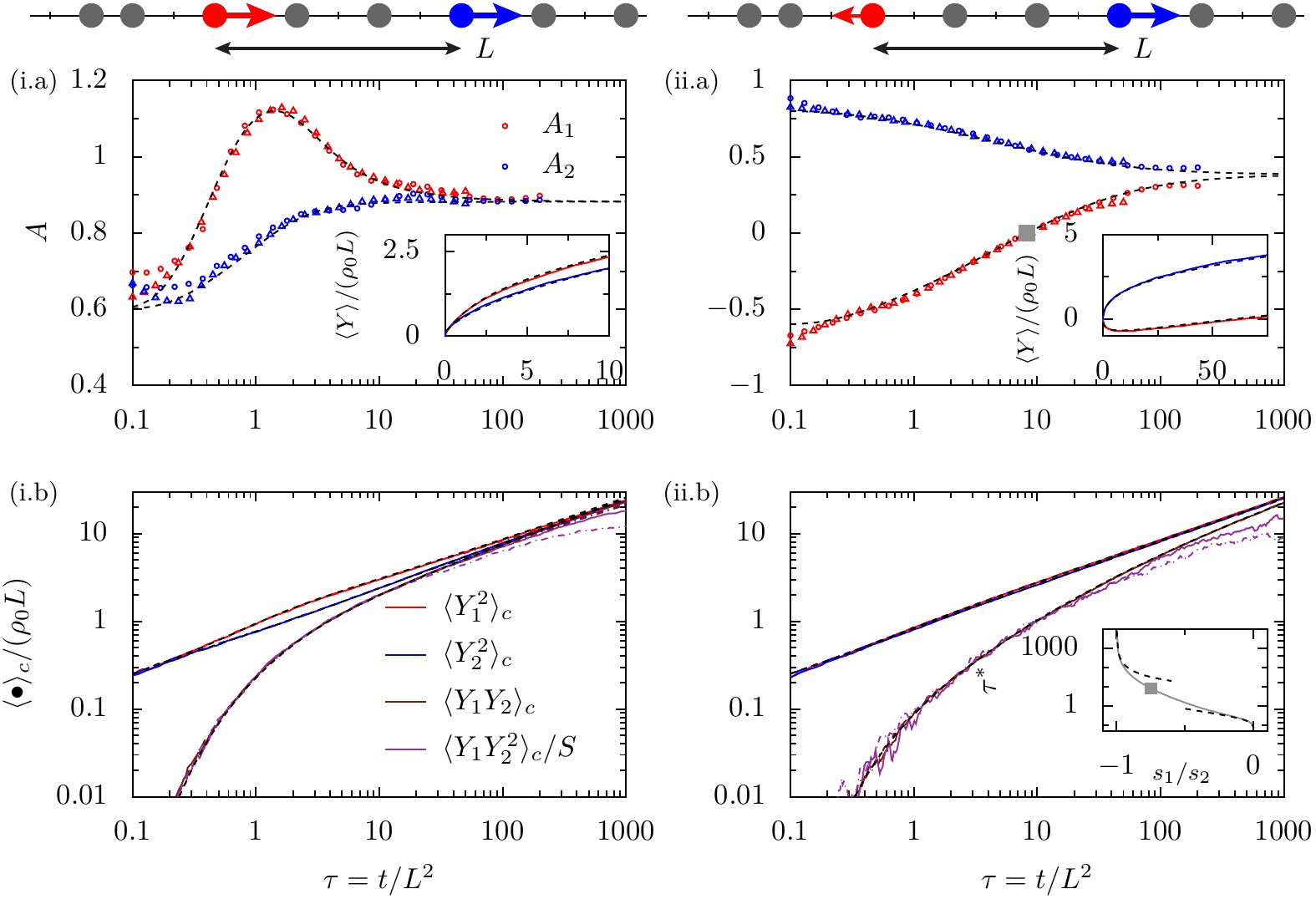}
\end{center}
\caption{Cooperativity and competition ($\rho_0 = 10^{-2}$).
(i) The two TPs have identical biases $s_1 = s_2 = 0.8$. (ii) The biases are in opposed directions $s_1 = -0.6, s_2=0.8$. The rescaled velocities $A$ are plotted on (i.a) and (ii.a) with the displacements in the inset, two initial distances are plotted: $L=50$ and $200$. At short time, the rescaled velocities are $s_1$ and $s_2$, at large time they are both equal to $S$.
The variances and some cumulants are plotted on (i.b) and (ii.b) (initial distances $L=10$ and $20$).
In the case (ii), the velocity of the first TP changes sign at rescaled time $\tau^\ast = t^\ast/L^2$ (gray square in ii.a).
We plotted the prediction $\tau^\ast$ as a function of $s_1/s_2$ for $s_2 = 0.8$. The dashed line are the asymptotic behaviors from Eq.~\eqref{eq:tast}, the grey square corresponds to the one in (ii.a).}
\label{fig:twoBiases}
\end{figure*}

{\it Bath-mediated entrainment.}
We examine the case of a single biased TP ($s_2\neq 0$) followed by an unbiased TP ($s_1 = 0$), initially separated by a distance $L$ (see Figure \ref{fig:oneBias}, top), which allows us to quantify the perturbation induced by a biased tracer in a quiescent medium. While the  behaviour of  $\langle Y_2\rangle$ is known
~\cite{Illien:2013}, we unveil an 
interesting scaling behaviour of $\langle Y_1\rangle$ beyond the large-time regime. Indeed, in the limit $t\to\infty$ with $t/L^2$ constant, one finds
\begin{gather}
 \frac{\langle Y_2(t)\rangle}{\rho_0} = s_2\sqrt\frac{2t}{\pi}, \qquad 
 \frac{\langle Y_1(t)\rangle}{\rho_0} = s_2\sqrt\frac{2t}{\pi} g\(\frac{L}{\sqrt{2t}}\),
 \label{eq:displ1bias} \\
g(u) = e^{-u^2} - \sqrt\pi\, u\, \text{erfc}(u).
\end{gather}
This provides the  dynamics of the entrainment of the TP1 by TP2, which admits a typical time scale $L^2$ and leads to the final  bound state 
discussed above 
(see Fig.~\ref{fig:oneBias}). 

The evolution towards the final regime can further  be quantified by  the dynamics of the two-TPs even (resp. odd) cumulants $\kappa^e = \langle Y_1^p Y_2^q\rangle_c$ (resp. $\kappa^o = \langle Y_1^p Y_2^q\rangle_c$) with $p+q$ even (resp. odd), $p,q\geq 1$.
They obey
\begin{equation} \label{eq:cums1bias}
 \frac{\kappa^e}{\rho_0} = \frac{\kappa^o}{\rho_0 s_2} = \sqrt\frac{2t}{\pi} g\(\frac{L}{\sqrt{2t}}\).
\end{equation}
Several comments are in order. 
(i) Equations ~\eqref{eq:displ1bias} and \eqref{eq:cums1bias}
are similar to the expressions found for the random average process~\cite{Rajesh:2001,Cividini:2016}, which points towards their universality.
(ii) The same scaling function $g$ is involved in the expressions of $\langle Y_1(t) \rangle$ and $\langle Y_1 Y_2(t)\rangle_c$ (Eqs~\eqref{eq:displ1bias} and \eqref{eq:cums1bias}).
This leads  to the generalized fluctuation-dissipation relation
\begin{equation}
\lim_{f_2\to0} \frac{2}{\beta}\frac{
 \left\langle Y_1\left(f_1=0, f_2\right)\right\rangle}{f_2} = 
\langle Y_1 Y_2\rangle _c(f_1=f_2=0).
\end{equation}
Note that  this relation holds in the opposite limit of a dilute ($\rho\to0$) SEP~\cite{Ooshida:2018}.
(iii)  Our approach provides the
time dependence of  all cumulants of individual particles and the law of the distance between TPs (insets of Fig.~\ref{fig:oneBias} and \cite{SupplMat}).
The time, initial distance between TPs, and driving force dependences from numerical simulations are unambiguously captured by our theoretical expressions (Fig.~\ref{fig:oneBias}).

{\it Bath-mediated cooperatively and competition.}
We now turn to the general case in which both TPs are biased (see Figure \ref{fig:twoBiases}, top).
The dynamics of effective interactions between TPs at the level of averages is conveniently analysed by introducing the rescaled instantaneous velocities
\begin{equation}
 A_j(t) = \frac{\sqrt{2\pi t}}{\rho_0} \frac{d\langle Y_j\rangle}{dt},
\end{equation}
which satisfy
$A_j(t)=s_j$ at small time and 
$\displaystyle A_j(t)= S=\frac{s_1 + s_2}{1+s_1s_2}$ at large time.
The full
time dependency is found~\cite{SupplMat} to be given by $A_1 = H_{s_1, s_2(1+s_1), -s_1s_2}(L/\sqrt{2t})$ and
$A_2 = H_{s_2, s_1(1-s_2), s_1s_2}(L/\sqrt{2t})$ with
\begin{equation} \label{eq:exprA}
 H_{\beta_0,\beta_1,\beta_2}(u) = \sum_{n=0}^\infty (-s_1s_2)^n \sum_{m=0}^2 
 \beta_m e^{-[(2n+m)u]^2}
\end{equation}
while higher order  cumulants follow
\begin{gather} \label{eq:cums2biases}
 \frac{\kappa^e}{\rho_0} = \frac{\kappa^o}{\rho_0 S} = \sqrt\frac{2t}{\pi} G_{s_1s_2}\(\frac{L}{\sqrt{2t}}\), \\
 G_\sigma(u) = (1+\sigma)\sum_{n=0}^\infty (-\sigma)^n g([2n+1]u).
\end{gather}

These results fully quantify the dynamics of the BMIs between two biased TPs and  reveal  striking behaviors. (i) In the case of same sign biases,
 the TPs cooperate~\cite{Benichou:2018} (Fig~\ref{fig:twoBiases} left).
At large times  a pair of biased TPs moves faster than a single TP-
in agreement with the asymptotic result Eq.~\eqref{eq:1tp}.  Note that such an accelerated 
dynamics has been numerically observed  in two-dimensional systems \cite{Mejia-Monasterio:2011,Vasilyev:2017}.
At intermediate times, we unveil an overshoot of the rescaled velocity of the trailing TP.
(ii)  When the biases act in opposite directions (say $0 < -s_1 < s_2$), each TP starts to move in the direction of its own force, and eventually both TPs move in the direction of the largest force (Figure~\ref{fig:twoBiases} left).
This  competing stage can be quantified from Eq.~\eqref{eq:exprA} by determining the U-turn time $t^\ast$ at which the velocity of TP1 changes its sign.
This time $t^\ast$ vanishes when $s_1$ is small, and diverges when $s_1$ is close to $-s_2$
according to the scaling laws
\begin{align}
\frac{ t^\ast}{L^2} &\equi{s_1\to 0} \frac{1}{2\log(-s_2/s_1)}; &
\frac{ t^\ast }{L^2}&\equi{s_1\to -s_2} \frac{\gamma}{1+s_1/s_2}
\label{eq:tast}
\end{align}
with $\gamma = 2(1+s_2)^2/(1-s_2)$.
Figure~\ref{fig:twoBiases} shows an excellent quantitative agreement between the analytical predictions and the numerical simulations.

\begin{figure}
\begin{center}
\includegraphics{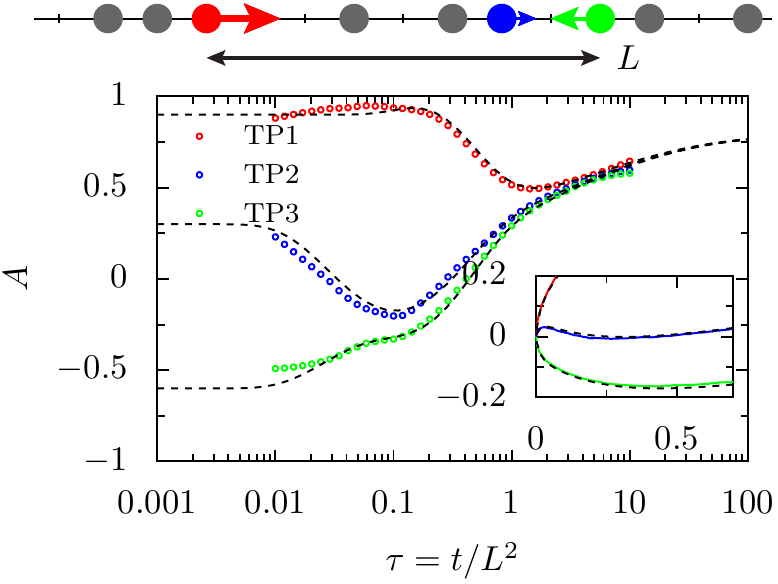}
\end{center}
\caption{Rescaled velocity of three TPs for biases $s_1 = 0.9, s_2 = 0.3, s_3 = -0.6$, total distance $L=X_3^0-X_1^0=60$ with $X_2^0 - X_1^0 = 45$, $\rho_0 = 10^{-2}$. The colored circles correspond to the numerical simulations while the dashed lines are the theoretical predictions~\cite{SupplMat}. Inset: average displacements of the TPs in linear rescaled time.
The behavior of the second TP displays several regimes: it first moves to the right, then under the influence of the third TP it goes to the left and finally the first TP pushed it back to the right.
Note that this complex dynamics is captured by our theoretical approach.}
\label{fig:threeBiases}
\end{figure}


Our approach can be extended to determine the dynamics of correlations in the case of an arbitrary number of driven TPs. Cooperativity and competition involve a complex cascade of  time scales associated with the initial distances between TPs, fully captured by our approach, as exemplified in the case of 3 TPs in Figure \ref{fig:threeBiases} and \cite{SupplMat}. 

{\it Bath-mediated interactions.}
The BMIs between two bound biased TPs can
be further analysed by associating the probability distribution of the
variation of distance $D=Y_2-Y_1$ to an effective potential $U(D)$ via $P(D)=\exp(-
\beta U(D))$~\cite{grosberg}.
For identical forces ($f_1 = f_2 = f$) and $D$ sufficiently close to its average value,
the two TPs are effectively bound by an harmonic potential
\begin{gather}
\label{D}
U(D) \sim \frac{\kappa}{2}\left(D - \left<D\right>\right)^2 \,,
\frac{\left<D\right>}{\rho_0 L} =  \left(\frac{1}{\cosh\left(\beta f\right)}-1\right) \,,
\end{gather}
where the constant $\kappa$ is explicitly given by
\begin{gather}
\label{kappa}
\kappa = \frac{\cosh\left(\beta f\right)}{\beta \rho_0 L \left(1+ \cosh\left(\beta f\right)\right)} \,.
\end{gather}
In the regime, $D \gg  \left<D\right>$, the potential displays a weaker dependence
on the distance $U(D) \sim D(\ln D + \nu)/\beta$ (see~\cite{SupplMat} for the value of $\nu$).
We remark that this qualitative change of regimes has been
observed in 2d in numerical simulations
of two biased TPs in a quiescent colloidal bath~\cite{Vasilyev:2017}.

Altogether, we determined the full dynamics of correlation functions
in a paradigmatic model of non equilibrium statistical physics
and entirely characterized the corresponding bath-mediated interactions.

\clearpage

\appendix

\renewcommand{\theequation}{S\arabic{equation}}
\renewcommand{\thefigure}{S\arabic{figure}}
\setcounter{figure}{0}

\begin{widetext}
 
\section{Details of the calculations}
\subsection{Model and approximation}
\subsubsection{Model}
Our model is based on the well known Symetric Exclusion Principle (SEP). Let us consider particles on a discrete one-dimensional line (the sites correspond to the integers).
The mean density is denoted $\rho \in [0, 1]$.
The particles follow a symmetric random walk with hard-core exclusions in continuous time.
The rate of jumps of these particles to the left / right is 1/2.
In addition to these particles,
we consider two biased tagged particles (TPs), initially at positions $X_1^0 = 0$ and $X_2^0 = L$. They can jump to 
the right with probability $p_j = (1+s_j) / 2$ and to the left with probability
$p_{-j} = (1-s_j)/2$, where $s_1$ and $s_2$ are the biases on the two TPs.
See Fig.~1 on the article for a sketch of the system.

For simplicity reasons, we choose to present only the computations for two TPs.
The treatement of three or more TPs is very similar but it leads to heavier computations. Results for three TPs will be presented in the next section.

\subsubsection{Approximation and consequences}
We focus on the limit of a dense system ($\rho\to 1$) and we follow the evolution of the vacancies
as was done in Ref.~\cite{Illien:2013}. In this limit of small number of vacancies,
one can approximate the motion of the TPs as being
generated by the vacancies interacting independantly (i.e. not simultaneously) with them:
in the large density limit the events
corresponding to two vacancies interacting simultaneously with the TPs happen only with negligible probability.

Let us consider a system of size $N$ with $M$ vacancies and denote by $\vec Y(t) = (X_1(t)-X_1^0, X_2(t)-X_2^0)$
the vector of the displacements of the two tracers.
The probability $P^{(t)} (\vec Y|\{Z_j\})$ of having displacements $\vec Y$ at
time $t$ knowning that the $M$ vacancies started at sites $Z_1\dots Z_M$ is exactly given by~:
\begin{equation}
 P^{(t)} (\vec Y|\{Z_j\}) =
  \sum_{\vec Y_1,\dots,\vec Y_M} \delta_{\vec Y, \vec Y_1+\dots+\vec Y_M}
  \mathcal{P}^{(t)} (\{\vec Y_j\}|\{Z_j\})
\end{equation}
where $\mathcal{P}^{(t)} (\{\vec Y_j\}|\{Z_j\})$ is the probability of displacements $\vec Y_j$ due to the vacancy $j$,
for all $j$, knowing the initial positions of all the vacancies.

Making the approximation that we described above, we can
link it to the probability $p^{(t)}_Z(\vec Y)$ that the tracers have moved by $\vec Y$ at time $t$
due to a single vacancy that was initially at site $Z$:
\begin{equation}
 \mathcal{P}^{(t)} (\{\vec Y_j\}|\{Z_j\}) \underset{\rho\to 1}{\sim} 
  \prod_{j=1}^M p^{(t)}_{Z_j}(\vec Y_j)
\end{equation}
We then have~:
\begin{equation}\label{eq:approx}
 P^{(t)} (\vec Y|\{Z_j\}) \underset{\rho\to 1}{\sim}
  \sum_{\vec Y_1,\dots,\vec Y_M}
  \delta_{\vec Y, \vec Y_1+\dots+\vec Y_M}
  \prod_{j=1}^M p^{(t)}_{Z_j}(\vec Y_j)
\end{equation}

We take the Fourier transform and we average over the initial positions of the vacancies:
\begin{equation}
\tilde p^{(t)}(\vec k) \equiv \frac{1}{N-2} \sum_{Z\neq 0,L} \sum_{\vec Y} p^{(t)}_Z(\vec Y) e^{i\vec k\cdot\vec Y}
\end{equation}
and mutatis mutandis for $\tilde P^{(t)} (\vec k)$. We obtain
\begin{equation}
 \tilde P^{(t)} (\vec k) \underset{\rho\to 1}{\sim} \left[\tilde p^{(t)}(\vec k)\right]^M
 = \left[\frac{1}{N-2}\sum_{Z\neq 0,L}\tilde p_Z^{(t)}(\vec k)\right]^M.
\end{equation}
Furthermore we write $\tilde p_Z^{(t)}(\vec k) = 1 + \tilde q_Z^{(t)}(\vec k)$ ($q_Z$ corresponds to the deviation
from a Dirac centered in $0$) to get
\begin{equation} \label{eq:fourier1}
 \tilde P^{(t)} (\vec k) \underset{\rho\to 1}{\sim}
 \left[1 + \frac{1}{N-2}\sum_{Z\neq 0,L} \tilde q_Z^{(t)}(\vec k)\right]^M.
\end{equation}

We now take the limit $N,M\to\infty$ with $\rho_0 \equiv 1-\rho = M / N$ (density of vacancies) remaining constant.
The second characteristic function reads
\begin{equation} \label{eq:psi1}
 \psi^{(t)}(\vec k) \equiv \ln \left[\tilde P^{(t)}(\vec k)\right] \approx \rho_0 \sum_{Z \neq 0,L} \tilde q_Z^{(t)}(\vec k).
\end{equation}
$\psi$ gives all the cumulants~: the cumulant associated to $Y_1^qY_2^r$ is 
$\langle Y_1^q Y_2^r\rangle = i^{-(q+r)} \left. \frac{\partial^{q+r}}{\partial k_1^qk_2^r}\psi \right|_{\vec k = \vec 0}$.

Our goal is now to find an expresion for $\psi^{(t)}(\vec k)$.

\subsection{Resolution}
The approximation (\ref{eq:approx}) that leads to (\ref{eq:fourier1}) tells us that the case
of a vanishing density of vacancies can be deduced from the case of a single vacancy. We first
focus on the latter before using (\ref{eq:psi1}) to get the cumulants.

\subsubsection{Single vacancy}
We consider the situation with a single vacancy. 
In the following we shall denote by $\nu = -1, +1, -2, +2$ the ``special sites'' respectively
to the left of TP1, to the right of TP1, to the left of TP2, to the right of TP2.
We use the vectors $\vec e_{\pm 1} = (\pm 1, 0)$ and $\vec e_{\pm 2} = (0, \pm 1)$.

The key is to introduce (conditional) first-passage time quantities
$F^{(t)}_{\eta, A}$
denotes the probability that the vacancy that started from site $A$ ($A$ can be a special site) at time
$0$ arrives \emph{for the first time} to the position of one of the tracers at time $t$ conditionned on the fact that it
was on the special site $\eta$ at time $t-1$.

A subtelty is that this probability shall depend
on a quantity called $L^\ast = L^\ast_{\eta, Z}$ which is the distance between the tracers when a vacancy \emph{can}
arrive at site $\eta$ when it was at site $Z$ at the very begining.
Let us denote $z = \mathbb{I}(0 < Z < L)$, $Z$ being the inital position of the vacancy.
Example: we consider $\eta=+1$, the vacancy
\emph{has to be} between the two tracers to get to this special site before any other.
Now if $z=0$ (vacancy started outside of the tracers) the distance between the tracers is $L^\ast = L+1$
(if $z=1$, it would be $L^\ast = L$). In any case,
we always have $L^\ast = L + \alpha(z,\eta)$, $\alpha\in \{-1,0,1\}$.
In the following $F^{(j), z}_{\eta, A} \equiv F^{(j)}_{\eta, A}(L + \alpha(\eta, z))$.
Note that if $Z$ is the initial position of the vacancy, there is no ambiguity:
$F^{(j)}_{\eta, Z} = F^{(j)}_{\eta, Z}(L^\ast = L)$.

One can partition over the first passage of the vacancy
to the site of one of the tracers to get an expression for $\tilde q_Z$ which is the main quantity involved
in (\ref{eq:psi1}):
\begin{align}
p_Z^{(t)}(\vec Y) &= \delta_{\vec Y, \vec 0}  \left(1 - \sum_{j=0}^t \sum_{\nu = \pm 1, \pm 2} F^{(j)}_{\nu, Z}\right)
+ \sum_{j=0}^t \sum_{\nu = \pm 1, \pm 2} p_{-\nu}^{(t-j)}(\vec Y-\vec e_\nu) F^{(j)}_{\nu, Z} \\
\tilde p_Z^{(t)}(\vec k) &= 1 - \sum_{j=0}^t \sum_{\nu = \pm 1, \pm 2}
F^{(j)}_{\nu, Z} + \sum_{j=0}^t \sum_{\nu = \pm 1, \pm 2}
\tilde p_{-\nu}^{(t-j),z} (\vec k) e^{i\vec k\cdot\vec e_\nu} F^{(j)}_{\nu, Z}, \\
\label{eq:qAll}
\tilde q_Z^{(t)}(\vec k) &= - \sum_{j=0}^t \sum_{\nu = \pm 1, \pm 2}
\left[1 - \left(1+\tilde q_{-\nu}^{(t-j), z} (\vec k)\right)  e^{i\vec k\cdot\vec e_\nu} \right] F^{(j)}_{\nu, Z}.
\end{align}

An exponant $z$ to a quantity means that this quantity is computed taking into account $z = \mathbb{I}(0 < Z < L)$.

We now need an expression for $\tilde q_\eta^z$ where $\eta$ is a special site.
To do so we decompose the propagator of the displacements over the successive passages of the vacancy to
the position of one of the tracers:
\begin{multline} \label{eq:decompo}
 p^{(t),z}_\eta (\vec Y) = \delta_{\vec Y,\vec 0} \left(1 - \sum_{j=0}^t \sum_\mu F^{(j),z}_{\mu, \eta}\right) \\
  + \sum_{p=1}^\infty \sum_{m_1,\dots,m_p=1}^{\infty} \sum_{m_{p+1}=0}^\infty
 \delta_{t, \sum_i\! m_i} \sum_{\nu_1, \dots, \nu_p}
  \delta_{\vec Y, \sum_i\! \vec e_{\nu_i}}
   \left(1- \sum_{j=0}^{m_{p+1}} \sum_\mu F^{(j),z}_{\mu, -\nu_p}\right)
  F^{(m_p),z}_{\nu_p, -\nu_{p-1}} \dots F^{(m_2),z}_{\nu_2, -\nu_1} F^{(m_1),z}_{\nu_1, \eta}
\end{multline}
the sums on $\mu$ and $\nu_i$ run over the special sites ($\pm 1, \pm 2$).

The discrete Laplace transform of a function of time $g(t)$ is
$\hat g(\xi)\equiv \sum_{t=0}^\infty g(t) \xi^t$.
We can now take both the Laplace and Fourier transforms of (\ref{eq:decompo}) to get:
\begin{equation}
 \hat p_{\eta}^z(\vec Y, \xi) = \frac{1}{1-\xi}\left\{ \delta_{\vec Y,\vec 0}
 \left(1 - \sum_\mu \hat F_{\mu,\nu}^z\right)
 + \sum_{p=1}^\infty \sum_{\nu_1, \dots, \nu_p} \delta_{\vec Y, \sum_i\! \vec e_{\nu_i}}
 \sum_\mu \left(1-\hat F_{\mu,-\nu_p}^z\right) \hat F_{\nu_p, -\nu_{p-1}}^z \dots \hat F_{\nu_2, -\nu_1}^z \hat  F_{\nu_1, \eta}^z 
 \right\}
\end{equation}

\begin{equation} \label{eq:qBorder}
 \hat{\tilde q}_\eta^{z} (\vec k, \xi)
 \equiv \hat{\tilde p}_\eta^z(\vec k, \xi) - \frac{1}{1-\xi}
 = \frac{1}{1-\xi} \sum_{\mu,\nu}
 \{[1-T^z(\vec k, \xi)]^{-1}\}_{\nu\mu} 
 \times \left(1-e^{-i\,\vec k\vec e_\nu}\right) e^{i\,\vec k\vec e_\mu}
 \hat F^z_{\mu\eta}(\xi)
\end{equation}
The matrix $T$ is defined by $T^z(\vec k,\xi)_{\nu\mu} = \hat F^z_{\nu,-\mu} (\xi) e^{i\,\vec k \vec e_{\nu}}$.

\subsubsection{Multiple vacancies}
Introducing (\ref{eq:qAll}) into (\ref{eq:psi1}), one gets an expression for the Laplace transform
of the second characteristic function:
\begin{align} \label{eq:psiSol}
 \hat \psi (\vec k, \xi) =& - \rho_0\bigg\{
 \sum_{\nu=-1,2} \left[
 \frac{1}{1-\xi} -  \left(\frac{1}{1-\xi} + \hat{\tilde q}_{-\nu}^{\;(z=0)} (\vec k, \xi)\right) e^{i\,\vec k\vec e_\nu} \right] h_\nu(\xi) \nonumber \\
 &+ \sum_{\nu=1,2} \left[
 \frac{1}{1-\xi} - \left(\frac{1}{1-\xi} + \hat{\tilde q}_{-\nu}^{\;(z=1)} (\vec k, \xi)\right) e^{i\,\vec k\vec e_\nu} \right] h_\nu(\xi)
 \bigg\} \\
 h_\nu(\xi) =& \sum_{Z\neq 0,L} \hat F_{\nu,Z} (\xi)
\end{align}
where $\hat{\tilde q}_{\nu}^z$ is given by (\ref{eq:qBorder}).

At the end of the day, the only quantities that need to be computed are~: $\hat F_{-1,-1}, \hat F_{2,2}$
(returning to the left of TP1 / to the right of TP2),
$\hat F_{1,1}$, $\hat F_{-2,-2}$ (returning to the left of TP1 without touching TP2, and vice-versa),
$F_{1,-2}, F_{-2,1}$ (arriving to the right of TP2 starting from the left of TP1 without returning, and vice-versa),
and the sums $h_{-1}, h_{2}$, $h_{1}, h_{-2}$.

\paragraph{Computation of the ``outside'' quantities}
We call ``outside'' quantities the ones related to the sites $-1$ and $2$: $\hat F_{-1,-1}(\xi)$, $\hat F_{2,2}(\xi)$,
$h_1(\xi)$ and $h_2(\xi)$. In this case, we do not care about the fact that there are two tracers: we simply consider
a random walk on a half line in which site $1$ is biased. Note that in this case, we do not care either about the distance
between the tracers (ie the value of $z$).

We first compute $F_{2,2}$: this amounts to the probability to return for the first time to the origin at time $t$
and being on site $1$ at time $t-1$ knowing that one was at site $1$ at time $0$.
We can partition on the first step of the walk.

For $\eta$ a ``special site'', we call $\tilde p_\eta = 1/(2p_\eta + 1)$ the probability for the vacancy to jump to the left ($\nu=1$) or right ($\nu=-1$) of a tracer when it is next to it.
\begin{equation}
 F_{2,2}(t) = \left(1-\tilde p_2\right) \delta_{t,1} + \tilde p_2 \sum_{j=1}^t f_1(j-1) F_{2,2}(t-j) \\
\end{equation}
where $f_l(t)$ the first-passage time density at the origin at time $t$ of a symmetric
Polya walk starting from site $l$.
One knowns $\hat f_l(\xi) = \alpha^{|l|}$ with $\alpha = \xi^{-1} (1-\sqrt{1-\xi^2})$~\cite{Hughes:1995}.

We take the Laplace transform and obtain:
\begin{equation} \label{eq:f22}
 \hat F_{2,2}(\xi) = \frac{(1-\tilde p_2)\xi}{1 - \tilde p_2\alpha\xi} 
\end{equation}
Similarly
\begin{equation} \label{eq:fm1m1}
 \hat F_{-1,-1}(\xi) = \frac{(1-\tilde p_{-1})\xi}{1 - \tilde p_{-1}\alpha\xi} 
\end{equation}

$h_{-1}$ and $h_2$ are easily computed from the former quantities:
\begin{align}
 F_{+2,Z}(t) &= \Theta(Z - (L+1)) \sum_{j=0}^t f_{Z-1}(j) F_{2,2}(t-j) \\
 \hat F_{+2,Z}(\xi) &= \Theta(Z - (L+1)) \hat f_{Z-L-1}(\xi) \hat F_{2,2}(\xi)
  = \Theta(Z - (L+1)) \hat F_{2,2}(\xi) \alpha^{Z-L-1} \\
 h_2(\xi) &= \sum_{Z\neq 0,L} \hat F_{+2,Z} (\xi) = \hat F_{2,2} \sum_{Z'=0}^\infty \alpha^{Z'}
 = \frac{\hat F_{2,2}(\xi)}{1-\alpha} \label{eq:h2} \\
 h_{-1}(\xi) &= \frac{\hat F_{-1,-1}(\xi)}{1-\alpha} \label{eq:hm1} 
\end{align}

\paragraph{Computation of the ``inside'' quantities}

We first solve the case with no bias $s_1 = s_2 = 0$. There are only two quantity to compute:
$\hat F^{(=)} \equiv \hat F_{+1,+1} = \hat F_{-2,-2}$ and $\hat F^{(\neq)} \equiv \hat F_{+1,-2} = \hat F_{-2,+1}$.
We also compute (for $0<Z<L$) $\hat F^\circ_Z \equiv \hat F_{+1, Z} = \hat F_{-2, L-Z}$.
($L$ is the distance between the tracers)

We recall the formula for the first passage time (of a Polya random walk) at site $s_1$ starting from $s_0$
and considering $s_2$ as an absorbing site~\cite{Hughes:1995},
\begin{equation}
 \hat f^\dagger (s_1|s_0,\xi) = \frac{\hat f_{s_1-s_0}(\xi) - \hat f_{s_1-s_2}(\xi) \hat f_{s_2-s_0}(\xi)}
 {1-\hat f_{s_1-s_2}(\xi)^2}
\end{equation}
with $\hat f_l(\xi) = \alpha^{|l|}$ in our 1-dimensional case. This gives us:
\begin{align}
 \hat F^{(=)}(\xi, L) &= \frac{\alpha - \alpha^{2L-1}}{1-\alpha^{2L}} \label{eq:Feq} \\
 \hat F^{(\neq)}(\xi, L) &= \frac{\alpha^{L-1} - \alpha^{L+1}}{1-\alpha^{2L}} \label{eq:Fneq} \\
 \hat F^\circ_Z(\xi, L) &= \frac{\alpha^{Z} - \alpha^{2L-Z}}{1-\alpha^{2L}} \label{eq:Fcirc}
\end{align}

The case $s_1 \neq 0, s_2 \neq 0$ can be deduced from the unbiased case for a distance $L-2$ by partioning on the
first passage of the vacancy to either +1 or -2:
\begin{align}
 F_{1,1}(t, L) =& (1-\tilde p_1) \delta_{t,1} \nonumber \\
 &+ \tilde p_1 \left( \sum_{j=1}^t F^{(=)} (j-1, L-2) F_{1,1} (t-j, L)
 + \sum_{j=1}^t F^{(\neq)} (j-1, L-2) F_{1,-2} (t-j, L) \right) \\
 \hat F_{1,1}(\xi, L) =& (1-\tilde p_1) \xi
 + \tilde p_1\xi \left( \hat F^{(=)} (L-2) \hat F_{1,-2} (L)
 + \hat F^{(\neq)} (L-2) \hat F_{1,-2} (L) \right)
\end{align}
The first term corresponds to a return to $1$, the second to a passage to $-2$.

This leads us to a system of $2\times 2$ equations with $2\times 2$ unknowns.
\begin{equation}
\left\{
   \begin{array}{l l l l}
      \left(\xi \tilde p_1 \hat F^{(=)}(L-2) - 1\right) \hat F_{1,1} &+ \left(\xi \tilde p_1 \hat F^{(\neq)}(L-2)\right) \hat F_{1,-2}
        &+ (1-\tilde p_1)\xi  & = 0 \\
      \left(\xi \tilde p_{-2} F^{(=)}(L-2) - 1)\right)\hat F_{1,-2} &+ \left(\xi \tilde p_{-2} F^{(\neq)}(L-2)\right) \hat F_{1,1} & & = 0 \\
      \left(\xi \tilde p_{-2} \hat F^{(\neq)}(L-2) - 1\right) \hat F_{-2,-2} &+ \left(\xi \tilde p_{-2} \hat F^{(\neq)}(L-2)\right) \hat F_{-2,1}
        &+ (1-\tilde p_{-2})\xi  & = 0 \\
      \left(\xi \tilde p_1 F^{(=)}(L-2) - 1)\right)\hat F_{-2,1} &+ \left(\xi \tilde p_1 F^{(\neq)}(L-2)\right) \hat F_{-2,-2} &  & = 0
   \end{array} \right .
\end{equation}
The solution is:
\begin{equation} \label{eq:Fint}
\left\{
   \begin{array}{r l}
   \hat F_{1,1} =& \displaystyle\frac{(1-\xi \tilde p_{-2}\hat F^{(=)})(1-\tilde p_1)\xi}
     {(1-\xi \tilde p_1 \hat F^{(=)})(1-\xi \tilde p_{-2} \hat F^{(=)}) - \xi^2 \tilde p_1 \tilde p_{-2} (\hat F^{(\neq)})^2} \\
   \hat F_{1,-2} =& \displaystyle\frac{\xi^2 \tilde p_{-2} (1-\tilde p_1) \hat F^{(\neq)}}
     {(1-\xi \tilde p_1 \hat F^{(=)})(1-\xi \tilde p_{-2} \hat F^{(=)}) - \xi^2 \tilde p_1 \tilde p_{-2} (\hat F^{(\neq)})^2}  \\
   \hat F_{-2,-2} =& \displaystyle\frac{(1-\xi \tilde p_1 \hat F^{(=)})(1-q_{-B})\xi}
     {(1-\xi \tilde p_1 \hat F^{(=)})(1-\xi \tilde p_{-2} \hat F^{(=)}) - \xi^2 \tilde p_1 \tilde p_{-2} (\hat F^{(\neq)})^2} \\
   \hat F_{-2, 1} =& \displaystyle\frac{\xi^2 \tilde p_1 (1-\tilde p_{-2}) \hat F^{(\neq)}}
     {(1-\xi \tilde p_1 \hat F^{(=)})(1-\xi \tilde p_{-2} \hat F^{(=)}) - \xi^2 \tilde p_1 \tilde p_{-2} (\hat F^{(\neq)})^2}  \\
   \end{array} \right .
\end{equation}
with $\hat F^{(=)}$ and $\hat F^{(\neq)}$ evaluted for $L-2$.

For $0 < Z < L$, one can again partion on the first passage of the vacancy to $+1$ or $-2$.
\begin{align}
 F_{+1,Z}(t, L) &= \sum_{j=1}^t F^\circ_{Z-1}(j, L-2) F_{1,1}(t-j, L) + \sum_{j=1}^t F^\circ_{L-Z-1}(j, L-2) F_{-2,-2}(t-j, L) \\
 \hat F_{+1,Z}(\xi, L) &= \hat F^\circ_{Z-1}(L-2) \hat F_{1,1}(L) + \hat F^\circ_{L-Z-1}(L-2) \hat F_{-2,-2}(L)
\end{align}

Using (\ref{eq:Fcirc}), we obtain:
\begin{align}
 h_1(\xi) = \sum_{Z=1}^{L-1} \hat F_{+1,Z} = \frac{(1-\alpha^{L-2})(1-\alpha^{L-1})}{1-\alpha^{2(L-2)}}
  \left(\hat F_{+1,+1} + \hat F_{+1,-2}\right) \label{eq:h1} \\
 h_{-2}(\xi) = \sum_{Z=1}^{L-1} \hat F_{-2,Z} = \frac{(1-\alpha^{L-2})(1-\alpha^{L-1})}{1-\alpha^{2(L-2)}}
  \left(\hat F_{-2,-2} + \hat F_{-2,+1}\right) \label{eq:hm2}
\end{align}

\subsection{Results}
\subsubsection{Symbolic calculation}
We use a symbolic calculation software (Mathematica) to put all the bricks together according to the
following graph. An asterix ($\ast$) indicates that one should be careful about which value of $L$
(distance between the TPs) should be used.

{\small
\begin{tikzpicture}
\node (psi) at (0,0.5)  {
\begin{tabular}{c}$\hat\psi(\vec k, \xi)$ \\ (\ref{eq:psiSol}) \end{tabular}};
\node (q) at (3,1) {
\begin{tabular}{c}$\hat q^z_\nu$ \\ (\ref{eq:qBorder})\end{tabular}};
\node (h) at (3,0) {
\begin{tabular}{c}$h_\nu$ \\ (\ref{eq:h2}, \ref{eq:hm1}, \ref{eq:h1}, \ref{eq:hm2})\end{tabular}};
\draw[->] (q) to[in=0,out=180] (psi);
\draw[->] (h) to[in=0,out=180] (psi);
\node (fext) at (7,1) {
\begin{tabular}{c}$\hat F_{-1,-1}$, $\hat F_{2,2}$ \\ (\ref{eq:f22}, \ref{eq:fm1m1})\end{tabular}};
\node (fint) at (7,0) {
\begin{tabular}{c}$\hat F_{\alpha,\beta}$ $(\alpha,\beta\in\{1,-2\})$ \\ (\ref{eq:Fint})\end{tabular}};
\draw[->] (fint) to[in=0,out=180] (q);
\draw[->] (fint) to[in=0,out=180] (h);
\draw[->] (fext) to[in=0,out=180] (q) node[above] {$\ast$};
\draw[->] (fext) to[in=0,out=180] (h);
\node (f0) at (10.5,0) {
\begin{tabular}{c}
  $\hat F^{(=)}$, $\hat F^{\neq}$ \\ (\ref{eq:Feq}, \ref{eq:Fneq})
\end{tabular}
};
\draw[->] (f0) to[in=0,out=180] (fint) node[above] {$\ast$};
\node (alpha) at (13.5,1) {$\alpha = \frac{1-\sqrt{1-\xi^2}}{\xi}$};
\node (p) at (13.5,0) {$\tilde p_\nu = \frac{1}{2p_\nu + 1}$};
\draw[->] (alpha) to[in=0,out=180] (fext);
\draw[->] (alpha) to[in=0,out=180] (f0);
\draw[->] (p) to[in=0,out=170] (fext);
\draw[->] (p) to[in=0,out=180] (f0);
\end{tikzpicture}
}

\subsubsection{Scaling limit}
We define a continuous Laplace variable $p=1-\xi$.
When $p$ goes to zero, $\alpha = (1-\sqrt{1-\xi^2})/\xi = 1 - \sqrt{2p} + \mathcal{O}(p)$.

We will be interested in the limit of large time $t$ and large distance $L$,
with $\tau = t/L^2$ constant. This amounts to writing $p = \tilde p / L^2$ with $L\to\infty$ and $\tilde p$ constant.
The inverse Laplace transform of a function $\hat f(p, L) = L^2 \hat g(\tilde p)$ is
$f(t, L) = g(\tau)$ with $g$ the inverse Laplace transform of $\hat g$.

Finally, our scaling limit in Laplace space amounts to the
following substitutions that can be performed with a numerical software (in this order).
\begin{align}
 1-\xi &\mapsto \tilde p/ L^2 &
 1-\alpha &\mapsto \sqrt{2\tilde p} / L &
 \alpha^L \approx \left(1 - \sqrt{2\tilde p} / L \right)^L  &\mapsto e^{-\sqrt{2\tilde p}} &
 \alpha \mapsto 1
\end{align}

\section{Detailed results}
\subsection{One TP}
We start by recalling the results of Ref.~\cite{Illien:2013} for a single TP of bias $s$ in the limit of high density (fraction of vacancies $\rho_0 \to 0$), at large time (small $p$).
\begin{align}
 \lim_{\rho_0\to 0} \frac{\hat\psi(k, p)}{\rho_0}
 &= \frac{1}{\sqrt{2} p^{3/2}}\left\{\cos k - 1 + s \sin k\right\} \\
 \lim_{\rho_0\to 0} \frac{\psi(k, t)}{\rho_0}
 &= \sqrt\frac{2t}{\pi}\left\{\cos k - 1 + s \sin k\right\}
\end{align}
The even (resp odd) cumulants of the displacement $Y$ are equal.
\begin{align}
 \frac{\langle Y^{2n}\rangle_c}{\rho_0} &=  \sqrt\frac{2t}{\pi} &
 \frac{\langle Y^{2n+1}\rangle_c}{\rho_0} &=  s\sqrt\frac{2t}{\pi}
\end{align}
The fraction $\langle\bullet\rangle$ is understood as a limit $\rho_0\to 0$.

\subsection{Two TPs}
We now focus on the case of two TPs with biases $s_1$ and $s_2$, initially separated by a distance $L$. The Laplace transform of the cumulant generating function reads
\begin{multline}
 \lim_{\rho_0\to 0} \frac{\hat\psi(\vec k, p=\tilde p / L^2)}{\rho_0}
 = \frac{L^3}{\sqrt{2} \tilde p^{3/2}}\bigg\{
 \hat K^{e,2}(\tilde p)(\cos (k_1+k_2) - 1) + \hat K^{o, 2}(\tilde p) \sin (k_1 + k_2) \\
 + \sum_{i=1}^2 \left[\hat K^{e,1}_i(\tilde p)(\cos k_i - 1) + \hat K^{o, 1}_i(\tilde p) \sin k_i\right]
 \bigg\}.
\end{multline}
with $\hat K^{\alpha, j}$ given in the following.
In real time, this gives
\begin{multline}
 \frac{1}{L}\lim_{\rho_0\to 0} \frac{\psi(\vec k, t=L^2\tau)}{\rho_0}
 = \sqrt\frac{2\tau}{\pi}\bigg\{
 K^{e,2}(\tau)(\cos (k_1+k_2) - 1) + K^{o, 2}(\tau) \sin (k_1 + k_2) \\
 + \sum_{i=1}^2 \left[K^{e,1}_i(\tau)(\cos k_i - 1) + K^{o, 1}_i(\tau) \sin k_i\right]
 \bigg\}.
 \label{eq:resPsi2TP}
\end{multline}
Note that $K(\tau)$ is related but not equal to the inverse Laplace transform of $\hat K(\tilde p)$.

From the previous expression one can deduce all the cumulants ($i=1,2$), including those
of the variation of distance $D=Y_2 - Y_1$.
\begin{align}
 \frac{\langle Y_1^{j}Y_2^{2n-j}\rangle_c}{\rho_0 L} &= K^{e,2}(\tau)\sqrt\frac{2\tau}{\pi} &
 \frac{\langle Y_1^{j}Y_2^{2n+1-j}\rangle_c}{\rho_0 L} &= K^{o,2}(\tau)\sqrt\frac{2\tau}{\pi}  \label{eq:cum2a} \\
 \frac{\langle Y_i^{2n}\rangle_c}{\rho_0 L} &= [K^{e,2}(\tau) + K^{e,1}_i(\tau)]\sqrt\frac{2\tau}{\pi} &
 \frac{\langle Y_i^{2n+1}\rangle_c}{\rho_0 L} &= [K^{o,2}(\tau) + K^{o,1}_i(\tau)]\sqrt\frac{2\tau}{\pi}  \label{eq:cum2b} \\
 \frac{\langle D^{2n}\rangle_c}{\rho_0 L} &= [K^{e,1}_1(\tau) + K^{e,1}_2(\tau)]\sqrt\frac{2\tau}{\pi} &
 \frac{\langle D^{2n+1}\rangle_c}{\rho_0 L} &= [K^{o,1}_2(\tau) - K^{o,1}_1(\tau)]\sqrt\frac{2\tau}{\pi}
\end{align}
Again, these results correspond to a limit $\rho_0\to 0$.

We now give the expressions of the quantities $\hat K(p)$ and $K(t)$ both in the cases
$s_1=0, s_2\neq 0$ and $s_1, s_2\neq 0$.

\[
\begin{array}{ccc}
 & \text{Case } s_1 = 0, s_2\neq 0 & \text{Case } s_1,s_2 \neq 0 \\[0.2cm]
 \tilde K^{e,2}(\tilde p)   & v & (1+s_1s_2)v/d_2 \\
 \tilde K^{o,2}(\tilde p)   & s_2 v & (s_1 + s_2)v/d_2 \\
 \tilde K^{e,1}_1(\tilde p) & (1-v)(1 + s_2 v) & (1-v)(1+s_2v)/d_2 \\
 \tilde K^{o,1}_1(\tilde p) & 0 & s_1(1-v)(1+s_2 v)/d_2 \\
 \tilde K^{e,1}_2(\tilde p) & 1-v & (1-v)(1-s_1 v)/d_2 \\
 \tilde K^{o,1}_2(\tilde p) & s_2 (1-v) & s_2 (1-v) (1-s_1 v)/d_2
\end{array}
\]
with $v = e^{-\sqrt{2\tilde p}}$ and $d_2 = 1+s_1s_2 v^2$.

\[
\begin{array}{ccc}
 & \text{Case } s_1 = 0, s_2\neq 0 & \text{Case } s_1,s_2 \neq 0 \\[0.2cm]
 K^{e,2}(\tau)   & g_1(\tau) & G_{0, s_1s_2, 0}(\tau) \\
 K^{o,2}(\tau)   & s_2 g_1(\tau) & G_{0, s_1+s2, 0}(\tau) \\
 K^{e,1}_1(\tau) & 1 + (s_2 - 1)g_1(\tau) + s_2 g_2(\tau) & G_{1,s_2-1,-s_2}(\tau) \\
 K^{o,1}_1(\tau) & 0 & G_{s_1,s_1(s_2-1),-s_1s_2}(\tau) \\
 K^{e,1}_2(\tau) & 1-g_1(\tau) & G_{1,-s_1-1,s_1}(\tau)  \\
 K^{o,1}_2(\tau) & s_2 (1-g_1(\tau)) & G_{s_2,-s_2(1+s_2),s_1s_2}(\tau)
\end{array}
\]
with
\begin{align}
 g(u) &= e^{-u^2} - \sqrt\pi u \erfc u \label{eq:gu}\\
 g_n(\tau) &= g(n/\sqrt{2\tau}) \\
 G_{\alpha,\beta,\gamma}(\tau) &= \sum_{n=0}^\infty
 (-s_1s_2)^n \left[\alpha g_{2n}(\tau) + \beta g_{2n+1}(\tau) + \gamma g_{2n+2}(\tau)\right]
\end{align}
Note that at small $\tau$, $G_{\alpha, \beta, \gamma} = \alpha$ while at
large $\tau$, $G_{\alpha, \beta, \gamma} = (\alpha + \beta + \gamma)/(1+s_1s_2)$.

The time evolution of the cumulants is given by Eqs.~\eqref{eq:cum2a}, \eqref{eq:cum2b}.
The rescaled velocities studied in the main text are obtained by taking the time derivative
of the average displacements.

\subsection{U-turn time for two TPs with opposite biases}
We consider two TPs with biases $s_2 > 0$ and $-s_2 < s_1 < 0$. We focus on the rescaled velocity of particle 1 defined as.
\begin{equation}
 A_1(t) \equiv \frac{\sqrt{2\pi t}}{\rho_0} \frac{d\langle Y_j\rangle}{dt}
\end{equation}

From~\eqref{eq:cum2b},
\begin{equation}
 \frac{\langle Y_1\rangle}{\rho_0 L} =
 \left[G_{0,s_1+s_2, 0}(\tau) + G_{s_1, s_1(s_2-1), -s_1s_2}(\tau)\right]
 \sqrt\frac{2\tau}{\pi}
 = G_{s_1, s_2(1+s_1), -s_1s_2}(\tau) \sqrt\frac{2\tau}{\pi}
\end{equation}
From this we deduce that
\begin{equation}
 A_1(\tau) = H_{s_1, s_2(1+s_1), -s_1s_2}(1/\sqrt{2\tau})
\end{equation}
\begin{equation} 
 H_{\beta_0,\beta_1,\beta_2}(u) = \sum_{n=0}^\infty (-s_1s_2)^n
 \left\{ \beta_0 e^{-[2nu]^2} + \beta_1 e^{-[(2n+1)u]^2} + \beta_2 e^{-[(2n+2)u]^2}\right\} \label{eq:H}
\end{equation}

At short time $A_1 = s_1 < 0$ while at large time $A_1 = (s_1+s_2)/(1+s_1s_2) > 0$.
There exists a rescaled time $\tau^\ast$ such that $A_1(\tau^\ast) = 0$.
One can solve numerically for $\tau^\ast$. Here we derive the asymptotic behavior
of $\tau^\ast$ when $s_1/s_2\to 0$ and when $s_1/s_2\to -1$.

When $s_1/s_2$ is small, $\tau^\ast$ is small so $(\tau^\ast)^{-1/2}$ is large.
We can keep only the first two terms in the sum \eqref{eq:H},
\begin{align}
 A_1(\tau^\ast) &\approx s_1 + s_2(1+s_1) e^{-\frac{1}{2\tau^\ast}} 
 \approx s_1 + s_2 e^{-\frac{1}{2\tau^\ast}} = 0, \\
 \tau^\ast &\approx \left[2\ln\left(\frac{s_2}{-s_1}\right)\right]^{-1}. \label{eq:tau1}
\end{align}

In the opposite limit, we write $s_1 = -s_2(1-\epsilon)$ with $\epsilon\ll 1$.
We guess that $(2\tau^\ast)^{-1} = \eta\epsilon$ with $\eta$ depending only on $s_2$.
The expansion gives
\begin{align}
 A_1 &= \sum_{n=0}^\infty (-s_1s_2)^n
 \left\{ s_1 (e^{-1/(2\tau)})^{(2n)^2} + s_2(1+s_1) (e^{-1/(2\tau)})^{(2n+1)^2}
 -s_1s_2 (e^{-1/(2\tau)})^{(2n+2)^2}\right\} \\
 &= \sum_{n=0}^\infty s_2^{2n} (1-n\epsilon) \times \nonumber \\
 &\left\{ -s_2(1-\epsilon) (1-(2n)^2\eta\epsilon) + s_2(1-s_2 + s_2\epsilon) (1-(2n+1)^2\eta\epsilon)
 +s_2^2(1-\epsilon) (1-(2n+2)^2\eta\epsilon)\right\} \\
 &= 0 + \epsilon\sum_{n=0}^\infty s_2^{2n} \left\{
 s_2(1+(2n)^2\eta) + s_2^2-s_2(1-s_2)(2n+1)^2\eta -s_2^2 -s_2^2(2n+2)^2\eta
 \right\} \\
 &= \epsilon\sum_{n=0}^\infty s_2^{2n+1} + \eta\epsilon
 \sum_{n=0}^\infty s_2^{2n}\left\{s_2(2n)^2 - s_2(1-s_2)(2n+1)^2 -s_2^2 (2n+2)^2\right\} \\
 &= \epsilon\left\{ \frac{s_2}{1-s_2^2} - \eta \frac{s_2(1+s_2)}{(1-s_2)^2} \right\}
\end{align}
$A_1 = 0$ gives
\begin{equation}
 \eta = \frac{s_2}{1-s_2^2} \frac{(1-s_2)^2}{s_2(1+s_2)}
 = \frac{1-s_2}{(1+s_2)^2}.
\end{equation}
At the end of the day,
\begin{equation}
 \tau^\ast \approx \frac{1}{2\eta\epsilon} = \frac{(1+s_2)^2}{2(1-s_2)} \frac{1}{1+\frac{s_1}{s_2}}. \label{eq:tau2}
\end{equation}

Eqs~\eqref{eq:tau1} and \eqref{eq:tau2} correspond to the asymptots of the inset
of Fig.~3, (b.ii) of the main text.

\subsection{Single trajectories}
We note that for large time and large distances, our results for cooperativity and competition
(Eqs. (4) and (9) of the main text)
actually hold
at the very level of a single trajectory. Numerical evidence is shown on Fig.~\ref{fig:sm:uniq}).

\begin{figure}
\begin{center}
\includegraphics{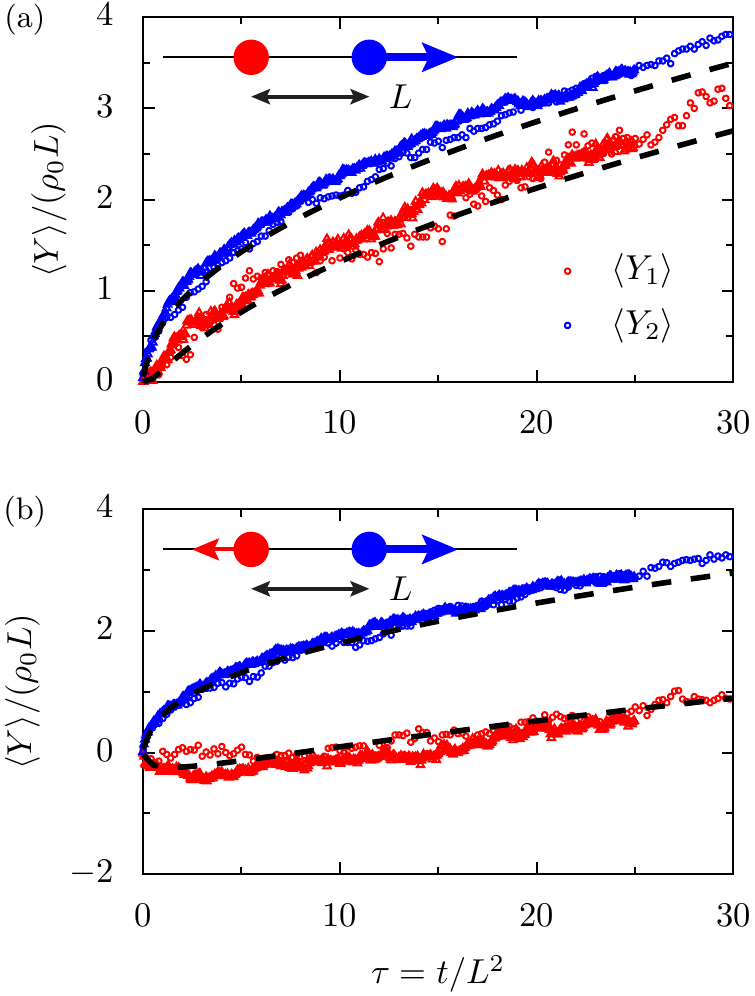}
\end{center}
\caption{Single realizations of the TP trajectories ($\rho_0 = 0.1$). (a) Entrainment: $s_1 = 0, s_2=0.8$. (b) Competition: $s_1 = -0.4$, $s_2=0.8$.
Circles: $L=1000$, triangles: $L=2000$, black lines: theoretical predictions from Eqs.~(4) and~(9) in the main text. We see that our approach is relevant
even for individual realizations.}
\label{fig:sm:uniq}
\end{figure}

\subsection{Three TPs}
We now turn to the case of three TPs with biases $s_1, s_2, s_3$. We denote
$L_1 = X_2^0 - X_1^0$ and $L_2 = X_3^0 - X_2^0$ and $L = L_1 + L_2$. Our result for the Laplace transform of the cumulant-generating function is
\begin{multline} \label{eq:res3tp}
 \lim_{\rho_0\to 0} \frac{\hat\psi(\vec k, p=\tilde p / L^2)}{\rho_0}
 = \frac{L^3}{\sqrt{2} \tilde p^{3/2}}\bigg\{
 \hat K^{e,3}(\tilde p)(\cos (k_1+k_2+k_3) - 1) + \hat K^{o, 3}(\tilde p) \sin (k_1 + k_2 + k_3) \\
 + \sum_{i=1}^2 \left[\hat K^{e,2}_{i, i+1}(\tilde p)(\cos (k_i + k_{i+1}) - 1) + \hat K^{o, 1}_i(\tilde p) \sin (k_i + k_{i+1})\right] \\
 + \sum_{i=1}^3 \left[\hat K^{e,1}_i(\tilde p)(\cos k_i - 1) + \hat K^{o, 1}_i(\tilde p) \sin k_i\right]
 \bigg\}.
\end{multline}
with
\begin{equation}
 K^{\alpha, n}(\tilde p) = \frac{\sum_{a, b=0}^2 Q^{\alpha, n}(a, b) v_1^a v_2^b}
 {1+s_1s_2v_1^2 + s_2s_3 v_2^2 + s_1s_3 v_1^2 v_2^2}
\end{equation}
with $v_1 = e^{-(L_1/L)\sqrt{2\tilde p}}, v_2 = e^{-(L_2/L)\sqrt{2\tilde p}}$.
For completeness, we give the 12 matrices $Q$ of coefficients.
{\small
\begin{align}
 Q^{e, 3} &= (1+s_1s_2+s_2s_3+s_1s_3)
 \begin{pmatrix}
   0 & 0 & 0 \\ 0 & 1 & 0 \\ 0 & 0 & 0  
 \end{pmatrix} &
 Q^{o, 3} &= (s_1+s_2+s_3+s_1s_2s_3)
 \begin{pmatrix}
   0 & 0 & 0 \\ 0 & 1 & 0 \\ 0 & 0 & 0  
 \end{pmatrix} \\
 Q^{e, 2}_{1,2} &= (1+s_1s_2)
 \begin{pmatrix}
   0 & 0 & 0 \\ 1 & s_3-1 & -s_3 \\ 0 & 0 & 0  
 \end{pmatrix} &
 Q^{o, 2}_{1,2} &= \frac{s_1+s_2}{1+s_1s_2} Q^{2, e}_{1,2} \\
 Q^{e, 2}_{2,3} &= (1+s_2s_3)
 \begin{pmatrix}
   0 & 1 & 0 \\ 0 & -1-s_1 & 0 \\ 0 & s_1 & 0  
 \end{pmatrix} &
 Q^{o, 2}_{2,3} &= \frac{s_2+s_3}{1+s_2s_3}Q^{2, e}_{2,3} \\
 Q^{e, 1}_1 &= 
 \begin{pmatrix}
   1 & 0 & s_2s_3 \\ s_2-1 & 0 & s_3(1-s_2) \\ -s_2 & 0 & -s_3  
 \end{pmatrix} &
 Q^{o, 1}_1 &= s_1 Q^{1, e}_1 \\
 Q^{e, 1}_2 &= 
 \begin{pmatrix}
   1 & s_3-1 & -s_3 \\ -1-s_1 & 1+s_1-s_3-s_1s_3 & s_3(1+s_1) \\ s_1 & s_1(s_3-1) & -s_1s_3  
 \end{pmatrix} & 
 Q^{o, 1}_2 &= s_2 Q^{1, e}_2 \\
 Q^{e, 1}_3 &= 
 \begin{pmatrix}
   1 & -1-s_2 & s_2 \\ 0 & 0 & 0 \\ s_1s_2 & -s1(1+s_2) & s_1  
 \end{pmatrix} &
 Q^{o, 1}_3 &= s_3 Q^{1, e}_3
\end{align}
}

The Laplace transform \eqref{eq:res3tp} can be inverted numerically to obtain the
time evolution of the cumulants.

One notices that the sum of all the coefficients of the matrices $Q^1$ and $Q^2$ is zero.
This means that for $\tilde p\to 0$, \eqref{eq:res3tp} simplifies into
\begin{equation}
 \lim_{\rho_0\to 0} \frac{\hat\psi(\vec k, p)}{\rho_0}
 = \frac{1}{\sqrt{2} \tilde p^3}\left\{
 \cos (k_1+k_2+k_3) - 1 + \frac{s_1+s_2+s_3+s_1s_2s_3}{1+s_1s_2 + s_1s_3 + s_2s_3} \sin (k_1 + k_2 + k_3)
 \right\}
\end{equation}
This corresponds to a single TP with effective bias $S = \frac{s_1+s_2+s_3+s_1s_2s_3}{1+s_1s_2 + s_1s_3 + s_2s_3}$. This behavior generalises to an arbitrary number of TPs,
the effective bias being
\begin{equation}
 S = \frac{\prod_{i=1}^N (1+s_i) - \prod_{i=1}^N (1-s_i)}{\prod_{i=1}^N (1+s_i) - \prod_{i=1}^N (1-s_i)} = \tanh\left(\frac{1}{2}\sum_{i=1}^N f_i\right)
\end{equation}
with $s_i = \tanh(f_i/2)$.

\subsection{Law of the distance between two TPs}
\subsubsection{Two TPs}
We denote $D = Y_2 - Y_1$ the \textit{variation} of distance between two TPs.
The cumulant-generating function of $D$ is given by
\begin{equation}
 \psi_D(q, t) \equiv \ln\left\langle e^{iq D(t)}\right\rangle 
 = \psi(k_1 = -q, k_2 = q, t)
\end{equation}
with $\psi$ given by Eq.~\eqref{eq:resPsi2TP}. Thus the Laplace transform of $\psi_D$ reads:
\begin{gather}
 \lim_{\rho_0\to 0} \frac{\hat\psi_D(q, p=\tilde p / L^2)}{\rho_0}
 = \frac{L^3}{\sqrt{2}\tilde p^{3/2}}\left\{
 \left[\hat K^{e,1}_1(\tilde p) + K^{e,2}_1(\tilde p)\right](\cos q - 1) + \left[\hat K^{o,1}_2(\tilde p) - \hat K^{o,1}_1(\tilde p)\right] \sin q 
 \right\} \\
 = \frac{L^3}{\sqrt{2}\tilde p^{3/2}}\frac{1-v}{1+s_1s_2 v^2} \left\{
 \left[2 + (s_2 - s_1)v\right](\cos q - 1)
 + i\left[s_2 - s_1 - 2s_1s_2 u\right]\sin q
 \right\}.
\end{gather}
At very-large time, ie when $\tilde p \to 0$, $v=e^{-\sqrt{2\tilde p}} \approx 1 - \sqrt{2\tilde p}$. This lead us to the following stationnary law $\psi_D^\text{stat}$ at large time:
\begin{align}
 \lim_{\rho_0\to 0} \frac{\hat\psi_D^\text{stat}(q, p)}{\rho_0}
 &= \frac{L}{p} \frac{\left[2 + s_2 - s_1\right](\cos q - 1)
 + i\left[s_2 - s_1 - 2s_1s_2 \right]\sin q}{1+s_1s_2} \\
 \lim_{\rho_0\to 0} \frac{\psi_D^\text{stat}(q)}{\rho_0}&= L \frac{\left[2 + s_2 - s_1\right](\cos q - 1)
 + i\left[s_2 - s_1 - 2s_1s_2 \right]\sin q}{1+s_1s_2} \\
 &= L \left\{ \frac{(1-s_1)(1+s_2)}{1+s_1s_2} (e^{iq} - 1) + (e^{-iq} - 1) \right\} \\
 &= L \left\{ \frac{2}{e^{\beta f_1} + e^{-\beta f_2}} (e^{iq} - 1) + (e^{-iq} - 1) \right\}. \label{eq:resLawDist}
\end{align}
We recall that the forces are defined by $e^{\beta f_i} = (1+s_i) / (1-s_i)$.
Eq.~\eqref{eq:resLawDist} corresponds to a Skellam distribution $P_D$ with parameters
$\mu_1 = 2\rho_0 L/(e^{\beta f_1} + e^{-\beta f_2})$ and $\mu_2 = \rho_0 L$.
\begin{equation}
 P_D(D) = e^{-(\mu_1 + \mu_2)}\left(\frac{\mu_1}{\mu_2}\right)^{k/2} I_k(2\sqrt{\mu_1\mu_2})
\end{equation}
The average of $D$ is
$\mu_1 - \mu_2$ and the variance is $\mu_1 + \mu_2$.

\subsubsection{Effective potential and Gaussian limit}
We focus on the case $f_1 = f_2 = f$: $\mu_1 = \rho_L[\cosh(\beta f)]^{-1}$, $\mu_2 = \rho_0 L$. In this case, we can associate an effective potential $U(D)$
to the probability law as
\begin{equation}
 P_D(D) \sim e^{-\beta U(D)}.
\end{equation}

For $D$ close to its average value $\mu_1-\mu_2$, the Skellam law is well approximated
by a Gaussian
\begin{equation} \label{eq:skellam}
 P_D(D) \approx \frac{1}{(\mu_1 + \mu_2)\sqrt{2\pi}} e^{-\frac{(D-\mu_1 + \mu_2)^2}{2(\mu_1 + \mu_2)}} 
 \sim e^{-\beta U(D)}
\end{equation}
with the effective potential $U(D)$ given by
\begin{align}
 U(D) &\approx  \frac{\kappa}{2}(D-\langle D\rangle)^2 \\
 \kappa &= \frac{1}{\beta(\mu_1 + \mu_2)} =
 \frac{\cosh (\beta f)}{\beta\rho_0 L[1 + \cosh (\beta f)]}.
\end{align}

\subsubsection{Large deviation function and large distance scaling}
Now, we want to derive the behavior of $U(D)$ at large $D$.
We first determine the large deviation function of the problem.
We have

\begin{align}
 P^\text{stat}_D(D) &= e^{-\rho_0 L (\tilde \mu_1 + 1)} \tilde\mu_1^{D/2} I_D(2\rho_0L\sqrt{\tilde \mu_1}), \\
 \ln P^\text{stat}_D(D) &= -\rho_0 L (\tilde \mu_1 + 1) + \frac{D}{2}\ln\tilde\mu_1
 + \ln I_D(2\rho_0L\sqrt{\tilde\mu_1})
\end{align}
with $\tilde\mu_1 = \mu_1/(\rho_0 L) = \cosh(\beta f)^{-1}$.
We use the expansion of the Bessel function at large order and large argument (here large $D \sim \rho_0 L$).
\begin{align}
 I_D(Dz) &\approx \frac{1}{\sqrt{2\pi D}} \frac{e^{\eta D}}{(1+z^2)^{1/4}} &
 \eta &= \sqrt{1+z^2}  + \ln \frac{z}{1+\sqrt{1+z^2}} \\
 \ln I_D(Dz) & \approx \eta D &
 \ln I_{2\rho_0 L x}(2\rho_0L x z) & \approx 2\rho_0 L x \eta &
\end{align}
Here $z = \sqrt{\tilde\mu_1} / x$. Thus
\begin{align}
 \eta = z\sqrt{1+\frac{1}{z^2}} - \ln\left(\frac{1}{z} + \sqrt{1+\frac{1}{z^2}}\right)
 =\frac{1}{x}\sqrt{x^2 + \tilde\mu_1} - \ln(x + \sqrt{x^2 + \tilde\mu_1}) + \frac{1}{2}\ln(\tilde\mu_1)
\end{align}
At the end of the day,
\begin{align}
 \frac{\ln P^\text{stat}_D(D=2\rho_0L x)}{2\rho_0 L}
 &= -\frac{\tilde\mu_1 + 1}{2} + \frac{x}{2}\ln\tilde\mu_1 + x\eta \\
 &= -\frac{\tilde\mu_1 + 1}{2} + \sqrt{x^2 + \tilde\mu_1} + x\ln\tilde\mu_1
 -x\ln(x + \sqrt{x^2 + \tilde\mu_1}) \\
 &\equiv \phi(x)
\end{align}
$\phi(x)$ is a large deviation function. The large $D$ behavior corresponds to the limit $x\to\infty$.
\begin{align}
 \phi(x) &\approx -x\ln x + (1-\ln 2 + \ln\tilde\mu_1) x, \\
 \ln P^\text{stat}_D(D) &\approx D\left(1 - \ln\frac{D}{\rho_0 L \tilde \mu_1}\right)
 \approx D\left(1 - \ln\frac{D \cosh(\beta f)}{\rho_0 L}\right) \\
 U(D) &= -\frac{1}{\beta}\ln P^\text{stat}_D(D)
 \approx \frac{D}{\beta}\left[\ln D + \ln\frac{\cosh(\beta f)}{\rho_0 L} -1\right] \equiv
 \frac{D}{\beta}\left[\ln D + \nu\right]
\end{align}
with $\nu = \ln\frac{\cosh(\beta f)}{\rho_0 L} -1$.

\subsubsection{Three TPs and non-additivity of the interactions}
We consider three TPs undergoing forces $f_1$, $f_2$ and $f_3$. We denote
$L_1 = X_2(t=0) - X_1(t=0)$ the initial distance between TPs $1$ and $2$.
Let us call $D_{12} = Y_2 - Y_1$ the variation of distance between TPs 1 and 2.
From~\eqref{eq:res3tp}, we show that its cumulant generating function at large time reads
\begin{align}
 \psi_{D_{12}}(q) &\equiv \ln\langle e^{iq D_{12}}\rangle \\
 \frac{\psi_{D_{12}}^\text{stat}(q)}{\rho_0}, &=
 L_1 \left\{\frac{2}{e^{\beta f_1} + e^{-\beta(f_2 + f_3)}}(e^{iq} - 1)
 + (e^{-iq} - 1)\right\}.
\end{align}
This still corresponds to a Skellam law.
The presence of a force $f_3$ modifies the stationnary law. We see that the interactions
between TPs are not additive.

Similarly, if we denote $L_2 = X_3(t=0) - X_2(t=0)$ and $D_{23} = Y_3 - Y_2$, we have
\begin{equation}
 \frac{\psi_{D_{23}}^\text{stat}(q)}{\rho_0} =
 L_2 \left\{\frac{2}{e^{\beta (f_1 + f_2)} + e^{-\beta f_3}}(e^{iq} - 1)
 + (e^{-iq} - 1)\right\}.
\end{equation}

\section{Numerical simulations}
\subsection{Simulations}
We do simulations of the SEP as follow. We consider a periodic line of $N$ sites.
We first place $n = 1, 2$ or $3$ tagged particles (TPs) at initial deterministic positions. 
Then, we place $M-n$ bath particles uniformly at random on the remaining sites.
For the figures of the main text we used $N=2500$, $M=2475$ ($\rho_0 = M / N = 0.01$).

At each time step we draw a particle at random and
try to move it either to the left or to the right according to its probability $p = (1+s)/2$ for the TPs,  $p = 1/2$ for the bath particles. The continuous time is
then incremented with a exponentially distributed random number with characteristic time $1/M$. This accounts for the exponentially distributed jump rates of the particles.
At equally spaced continuous times we compute the observables: the moments with $1$ to $n$ TPs.

We perform the average over multiple simulations to get good statistics. Typically we
perform $50000$ simulations (except for the Figure for a single realization).

\subsection{Numerical effective biases}
We now discuss how to obtain the numerical curves for the rescaled velocities defined as
\begin{equation}
 A(t) = \sqrt{2\pi t} \frac{dX}{dt}
\end{equation}
from the numerical data $X$ versus $t$. Let us define
\begin{equation}
 B(t) \equiv \sqrt\frac{\pi}{2t} X(t)
\end{equation}
which is easy to compute from the data. 
One sees that
\begin{equation}
 A(t) = B(t) + 2t\frac{dB}{dt}.
\end{equation}
Now, we write $w = \ln t$ which gives us
\begin{equation}
 A(w) = B(w) + 2 \frac{dB}{dw}.
\end{equation}

Our procedure is as follow: we first gather the data into bins of logarithmic size over
which we average. This gives us $B(w)$. We then compute the derivative
$dB/dw$ using a Savisky-Golay filter. We finally obtain $A(w)$ that we can plot on our graphs.

\end{widetext}

\end{document}